# Total transmission and total reflection by zero index materials

**Viet Cuong Nguyen\* and Lang Chen\*\***

50 Nanyang Avenue, School of Materials Science and Engineering
Nanyang Technological University, Singapore 639798, Singapore.

\*NG0067NG@e.ntu.edu.sg
\*\* Corresponding author: langchen@ntu.edu.sg

In this report, we achieved total transmission and reflection in a slab of zero index materials with defect(s). By controlling the defect's radius and dielectric constant, we can obtain total transmission and reflection of EM wave. The zero index materials, in this report, stand for materials with permittivity and permeability which are simultaneously equal to zero or so called matched impedance zero index materials. Along with theoretical calculations and simulation demonstrations, we also discuss about some possible applications for the proposed structure such as shielding or cloaking an object without restricting its view. We also suggest a way to control total transmission and reflection actively by using tunable refractive index materials such as liquid crystal and BST. The physics behind those phenomena is attributed to intrinsic properties of zero index materials: constant field inside zero index slab.

Metamaterials has been paid much attention recently because it can be applied to make several novel devices such as invisible cloak [1, 2], perfect lens [3], slow light devices [4] and much more interesting applications. It contains unusual engineered permittivity and permeability such as double negative index [5], single negative index [6], epsilon near zero and matched impedance zero index [7, 8]. Compared with double negative index and single negative index materials, Zero Index Metamaterials (ZIMs) has received much less attention, although its potential application may be as important as negative index materials.

ZIMs, whose permittivity and permeability are simultaneously or individually equal to zero, was investigated in details by several scientists [9]. Electromagnetic waves can be tunneled through very narrow channels filled with epsilon zero materials, which has been demonstrated experimentally at RF regime [10]. An example of epsilon zero materials is silicon carbide (SiC) in Infra red and optical frequency [11]. Matched impedance zero index materials (MIZIM), a sub branch of ZIMs, was investigated theoretically and proved to facilitate total transmission without changing phase and transform circular wave into plane wave [12]. The first man-made MIZIM was fabricated and it works in GHz range of frequency [8]. MIZIM is also made for mid-IR [13]. In mid-IR frequency, the index and impedance of fabricated MIZIM are reported to be -0.04-0.07j and 1.03-0.01j [13]. To avoid confusion, we note that materials with zero permittivity are called Epsilon Zero Materials (EZMs) (like suggested in [8]). Materials with permittivity and permeability are simultaneously equal to zero are called MIZIMs.

In this work, we will concentrate on MIZIMs with defect(s) as given in Fig 1. By controlling dielectric constant in region 2 (defect), we can obtain total reflection or total transmission of EM wave. We note that those effects can also happen in EZMs as well and it will be discussed along in the text. Moreover, we also discuss some possible applications of the structure such as invisible cloaking or shielding.

First, let us consider a structure shown in Fig 1. A 2-D transverse magnetic EM wave (TM) is excited at the far end of left hand side and received at the right hand side. Region 1 is MIZIMs and region 2 is a circular shape filled with normal materials ($\varepsilon_2$ and $\mu_2$). The electromagnetic wave in region 1 must satisfy the following Maxwell Eq.

$$E^{(1)} = \frac{1}{i\omega\varepsilon_1} \nabla \times H^{(1)} \tag{1}$$

where the superscriptions stand for field inside that region. As $\varepsilon_1$ in Eq. (1) tends to zero in a quasi-static case, hence $H^{(1)}$ must be constant in region (1) so that $E^{(1)}$ will be finite. Due to this condition, any scatterings from region 2 will result in a global increase or decrease in H field of region 1 rather than any local fluctuations. At boundary between region 2 and 1, $H^{(1)}$ must be equal with $H^{(2)}$ and $E^{(2)}$ must follow the below Eq..

$$E^{(2)} = \frac{1}{i\omega\varepsilon_2} \nabla \times H^{(2)} \tag{2}$$

On the other hand, EM wave in region 2 must follow Maxwell Eq. as

$$\nabla \times \nabla \times H^{(2)} + k^2 H^{(2)} = 0 \tag{3}$$

At region 0, the field is the sum of the incident field and the reflected field. Solutions of Maxwell Eq.s yield

$$H^{(0)} = H_0(\exp(ik_0 x) + R\exp(-ik_0 x)) \tag{4}$$

$$E^{(0)} = \frac{H_0}{i\omega\varepsilon_0}(-ik_0 \exp(ik_0 x) + ik_0 R\exp(-ik_0 x)) \tag{5}$$

R is reflection coefficient, $k_0$ is wave vector at free space, $\omega$ is angular frequency of the incident wave and $\mu_0$ is the permeability of free space. At region 3 the field is given as

$$E^{(3)} = H_0 \left( \frac{-ik_0}{i\omega\varepsilon_0} T \exp(ik_0(x-d)) \right) \tag{6}$$

$$H^{(3)} = H_0 (T \exp(ik_0(x-d))) \tag{7}$$

T is the transmission coefficient and d is the thickness of the MIZIMs slab. At x=0 and x=d, the tangential components of EM field must be continuous. Hence the relationship between EM field in region 1 and 0 is written as

$$(R+1)H_0 = H^{(1)} \tag{8}$$

The magnetic field inside region 1 is constant. Solving Eq. 3 with Dirichlet boundary condition, we obtain the following H-field inside region 2

$$H^{(2)} = H^{(1)} \frac{J_0(k_2 r)}{J_0(k_2 R)} \tag{9}$$

where $J_L$ is the Bessel function of the first kind and order l, $k_2$ is the wave vector inside region 2 and R is the radius of region 2. Electric field inside region 2 follows

$$E^{(2)} = -H^{(1)} \frac{k_2}{i\omega\varepsilon_2} \frac{J_1(k_2 r)}{J_0(k_2 R)} \tag{10}$$

Transmission through the structure is given in general form [14]

$$T = \frac{2d}{2d - \frac{Z_0}{H^{(1)}} \oint_{dA_2} \vec{E}^{(1)} \overrightarrow{dl}} \tag{11}$$

where $Z_0$ is the impedance of free space and d is the thickness of the slab. $dA_2$ denotes the path integral acts around region 2.

Carefully investigating of Eq. 8 and 9 indicates the possibility to obtain a perfect reflection for EM wave. At first, let us start with Eq. 9, when $k_2R$ is chosen such that denominator of (9) is zero, the H-field inside region 1 must be zero to maintain a finite value for $H^{(2)}$. Because H- field in region 1 is zero, the EM wave is total reflected as indicated in Eq. 8. There are several possible values of $k_2R$ to make $J_0(k_2R)$ equals to zero. The second possible way to get perfect reflection is: coat region 2 with PMC layers as suggested in [14]. However, this method eliminates field inside region 2. Following our method, EM wave is perfectly reflected while H field is still finite inside region 2. We note that this phenomenon can happen for both TM and TE wave.

Moreover, Eq. 10 and 11 suggest a possible way to get strong transmission of EM wave through the structure. By choosing the suitable value of $k_2R$, we can make $J_1(k_2R)$ equal to zero and hence $E^{(1)}$ at the boundary between region 1 and 2 will also be zero. From this point of argument, the path integral in Eq. 11 will be zero. Thus the transmission will be perfect. Moreover, this phenomenon happen for both TE and TM wave.

To demonstrate the idea, we will plot $J_0(k_2R)$ and $J_1(k_2R)$ versus index of region 2 ($n_2$). Figure 2a and 2b show such plots when R= 0.3 m and incident wavelength is 0.3 m. As indicated in these plots, there are several values of $n_2$ which make $J_0(k_2R)$ and $J_1(k_2R)$ equal to zero. When $J_0(k_2R)$ equals to zero, total reflection takes place. When $J_1(k_2R)$ equals to zero, perfect transmission holds.

To perform simulations on the proposed structures, Comsol Multiphysics with a FEM EM wave solver is used. At first, we demonstrate total reflection effect by simulation. Fig 3a shows H-field distribution in the simulation domain. In this figure, $\varepsilon_2 =$ 15 and R= 0.3 m, thickness of region 1 is arbitrary and incident wavelength is 0.3 m. We note that $\varepsilon_2$= 15 and R= 0.3 m are the suitable values to get zero $J_0(k_2R)$ and hence EM wave is totally reflected. The field inside region 2 is governed by Eq. 9 and 10. Fig 3b and 3c illustrate the total reflection effect for array of region 2 along y-axis and x-axis respectively. For both of these Figs, total reflection is still reserved. Fig 3d shows effect of perfect reflection when a portion of region 2 is coated with PMC layer. In Fig 3d, region 2 still has the following parameters: $\varepsilon_2$= 15 and R= 0.3 m.

On principle, total reflection effect can take place for other geometries different from circular shape. Fig 4a demonstrates total reflection effect for eclipse shape with $R_x$=0.125 m, $R_y$=0.1 m and $\varepsilon_2$= 12. Theoretically, we can obtain very high concentrated magnetic field inside region 2 by choosing suitable shape and dielectric constant for region 2. Fig 4b demonstrates a strong magnetic field concentrated inside region 2 with $R_x$=0.125, $R_y$=0.1 and $\varepsilon_2$= 54.4. The H-field inside region 2 can be 50 times larger than outside H- field. In brief summary, we note that when $J_0(k_2R)$ equals to zero, EM wave will be totally reflected. The perfect reflection phenomenon is governed by Eq. 8 and 9.

Secondly, we will demonstrate total transmission by simulation. As discussed in theory section, when $J_1(k_2R)$ equal to zero, there is possibility for total transmission of EM wave. Figure 5a shows plotting of transmission versus $\varepsilon_2$. The radius of region 2 is 0.3 m and incident wavelength is 0.3 m. From those plots, we see that each of total transmission peaks is corresponding to a point which makes $J_1(k_2R)$ equal to zero in Fig 2b. Each of total reflection dips is corresponding to a point which makes $J_0(k_2R)$ equals to zero in Fig 2a. Fig 5b shows distribution of EM field when perfect transmission condition is satisfied. In this Fig, EM field is transmitted perfectly without any scatterings. In brief summary for this paragraph, we note that when $J_1(k_2R)$ equals to zero, perfect transmission of EM wave is obtained. Perfect transmission phenomenon is governed by Eq. 10 and 11 and it works for both TE and TM wave.

In this paragraph, we discuss some possible applications. At first, we note that MIZIMs have been synthesized for IR frequency. This information makes the proposed structure easier to handle and to realize the effects. Secondly, MIZIMs will maintain constant EM field inside region 1 hence it makes EM wave transmit or reflect from the proposed structure without suffering from any scatterings. Hence, they suggest that the proposed structure can serve for cloaking or shielding technology. Due to finite field inside region 2, hided or cloaked object inside this region still can "sense" the surrounding. Region 2 will serve as an "observation windows" [2, 15]. The second possible application arises from Fig 5a. In this Fig, total transmission peaks and total reflection peaks are near to each other hence it suggests the possibility to control transmission and reflection state actively by tuning dielectric constant inside region 2. Tuning dielectric constant can be achieved by using liquid crystal or BST [16].

We note that all of the above phenomena can also take place when region 1 is EZMs instead of MIZIMs. However, when region 1 is EZMs, these phenomena can only work for TM incident wave. Moreover, these theories can be extended into optical range of frequency.

In summary, we demonstrated the possibility to obtain total transmission and total reflection in MIZIMs with dielectric defect(s). By controlling dielectric constant and radius of defect, we can achieve total transmission and total reflection of EM wave. The ideas are demonstrated in theory and modeling, simulations were also performed to confirm those theoretical calculations. In practice, MIZIMs have been synthesized for RF and mid-IR ranges which make the proposed structure more practical in real applications. We also discussed some possible applications of the proposed structure such as invisible cloaking or shielding an object without restricting its ability to "sense" the surrounding. We also suggest a method to control perfect transmission and perfect reflection actively by tuning dielectric constant of region 2.

**Acknowledgement**

The authors acknowledge the support from Nanyang Technological University and Ministry of Education of Singapore under Projects No. TL/SP/07-03, AcRF RG 21/07 and No. ARC 16/08.

**Figures (all color online) and captions:**

Fig 1: (a) Physical picture of the proposed vacuum/MIZIM/vacuum structure; region (0) and (3) are vacuum media while region (1) is MIZIM; region (2) is normal material.

Fig 2: Plotting of Bessel function of the first kind versus refractive index of region 2. (a) $J_0(k_2R)$ vs $n_2$. (b) $J_1(k_2R)$ vs $n_2$. When $J_0(k_2R)$ equals to zero, total reflection takes place. When $J_1(k_2R)$ equals to zero, perfect transmission holds. R= 0.3 m and incident wavelength is 0.3 m.

Fig 3: H-field distribution along simulation domain (a) single defect with R= 0.3 m and incident wavelength is 0.3 m, reflection is unity and wave reflected smoothly, thickness of region 1 is arbitrary. (b) Array of region 2 along y-axis. (c) Array of region 2 along x-axis. (d) A small portion of region 2 is coated with PMC; EM wave is still reflected while field inside region 2 is finite.

Fig 4: (a) H-field distribution when region 2 is an eclipse with $R_x$=0.125 m, $R_y$=0.1 m and $\varepsilon_2$= 12. Total reflection is reserved. (b) $R_y$=0.1 m and $\varepsilon_2$= 54.4, H-field inside region 2 is 50 times stronger than outside H-field. Incident wavelength is still 0.3 m.

Fig 5: (a) Transmission versus permittivity of region 2 when R= 0.3 m and incident wavelength is 0.3 m. (b) H- field distribution when perfect transmission condition is satisfied.

**Fig.1**

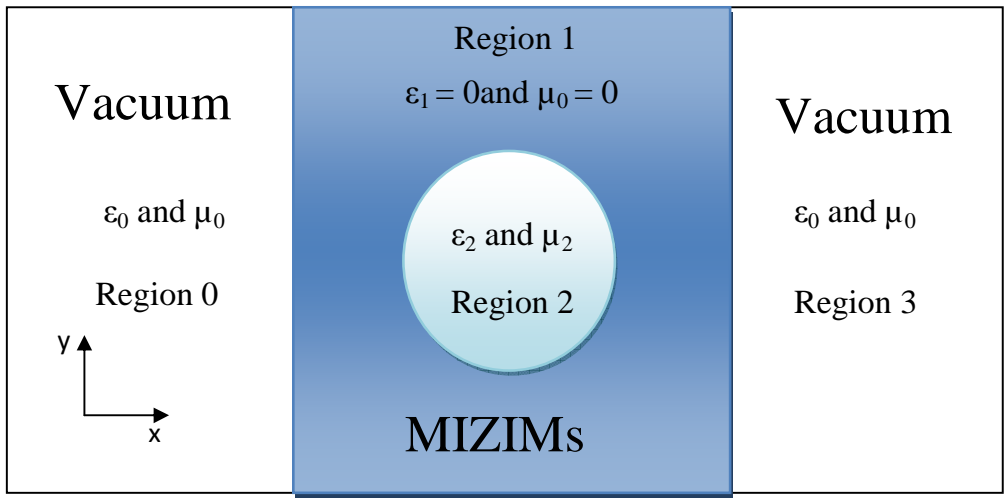

**Fig 2**

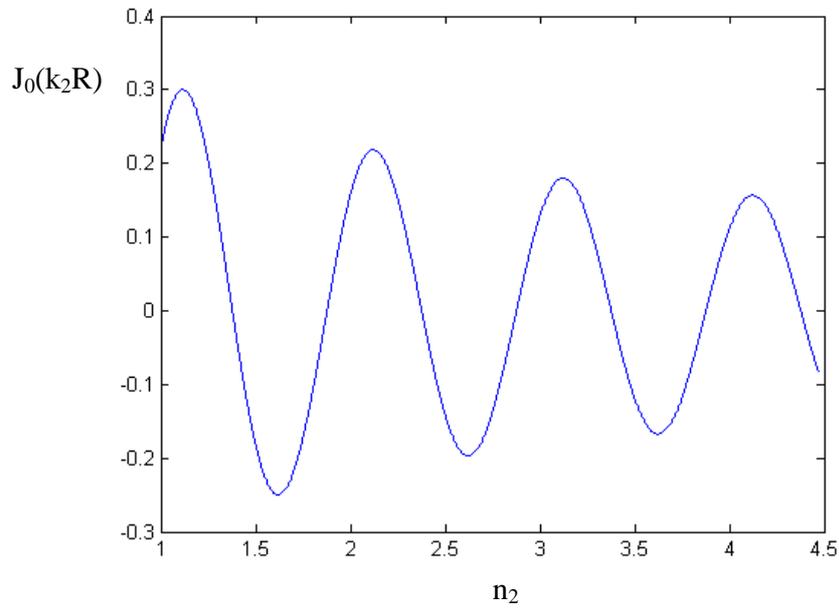

(a)

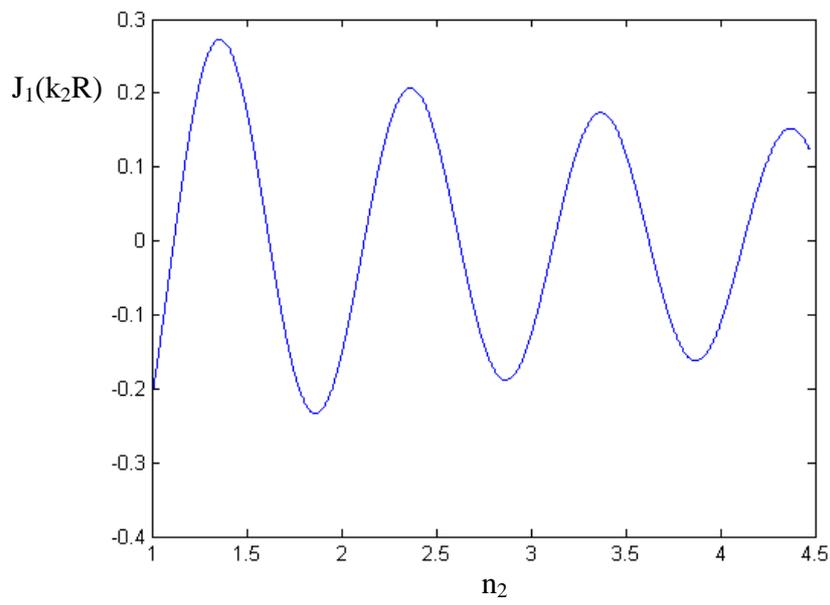

(b)

**Fig. 3**

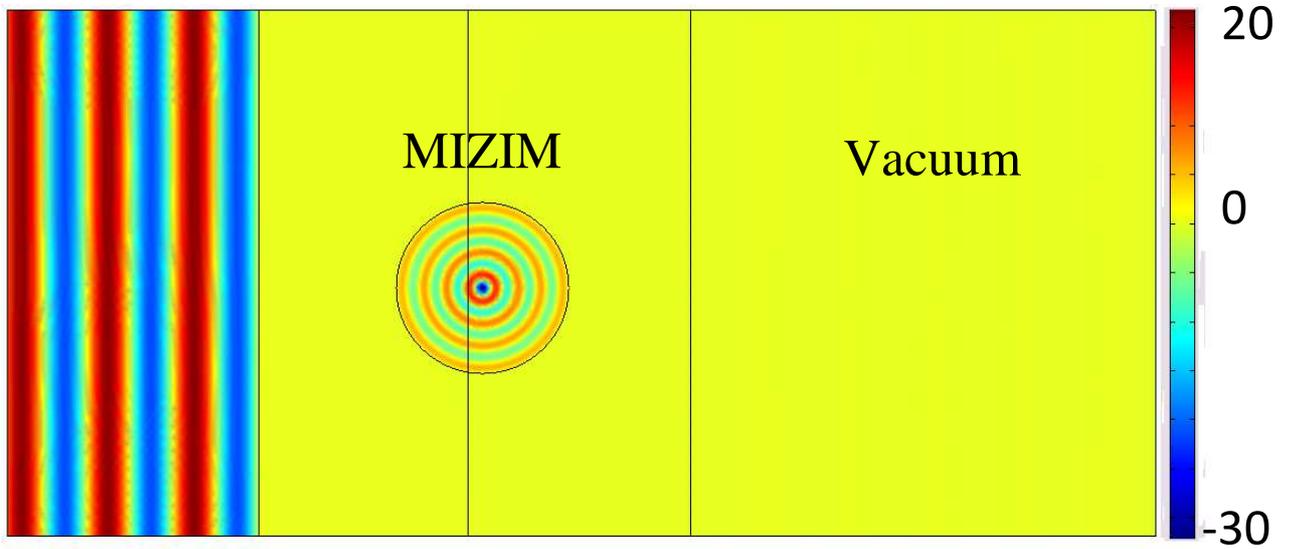

(a)

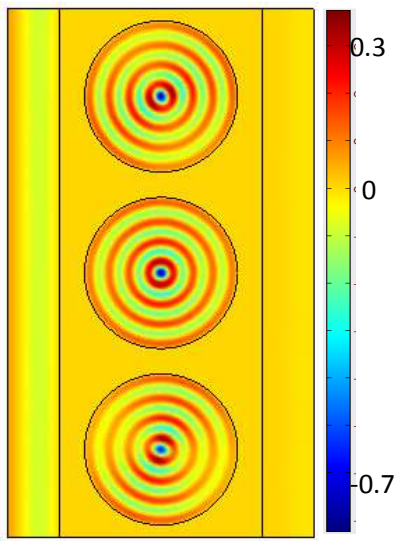

(b)

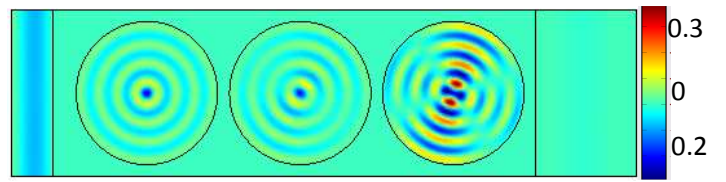

(c)

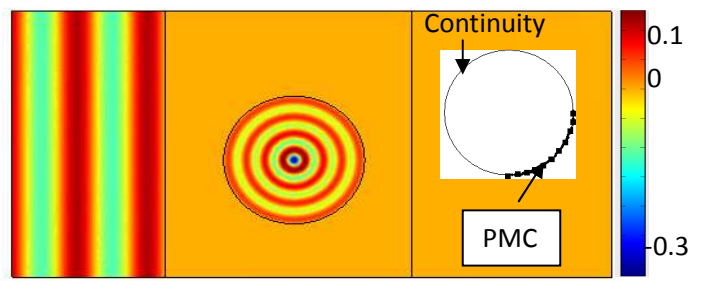

(d)

13  Nguyen et al

**Fig. 4**

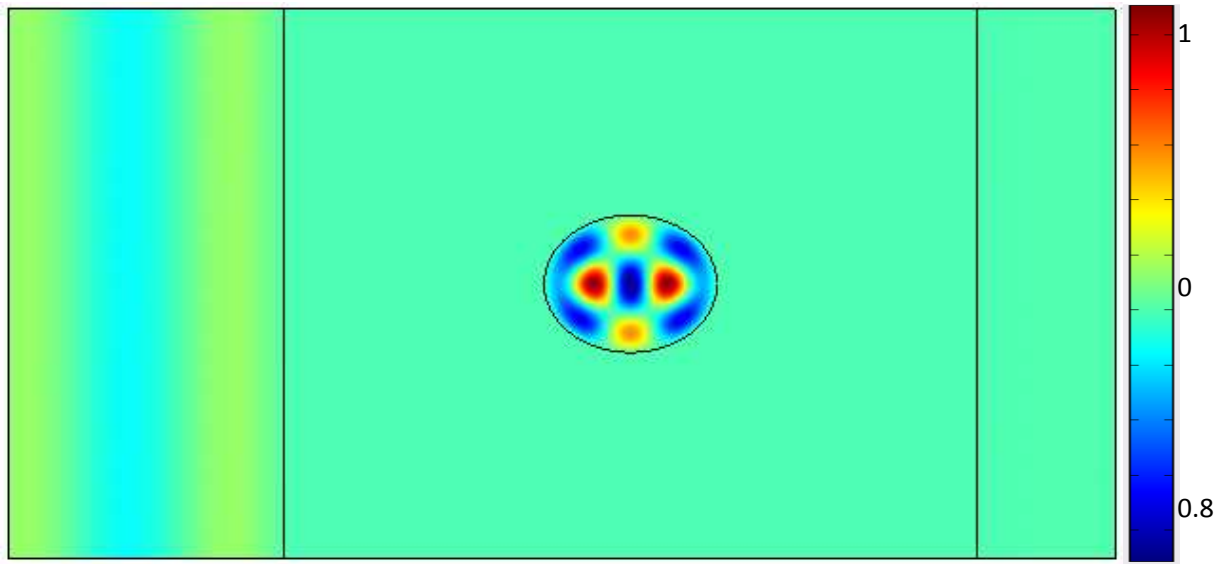

(a)

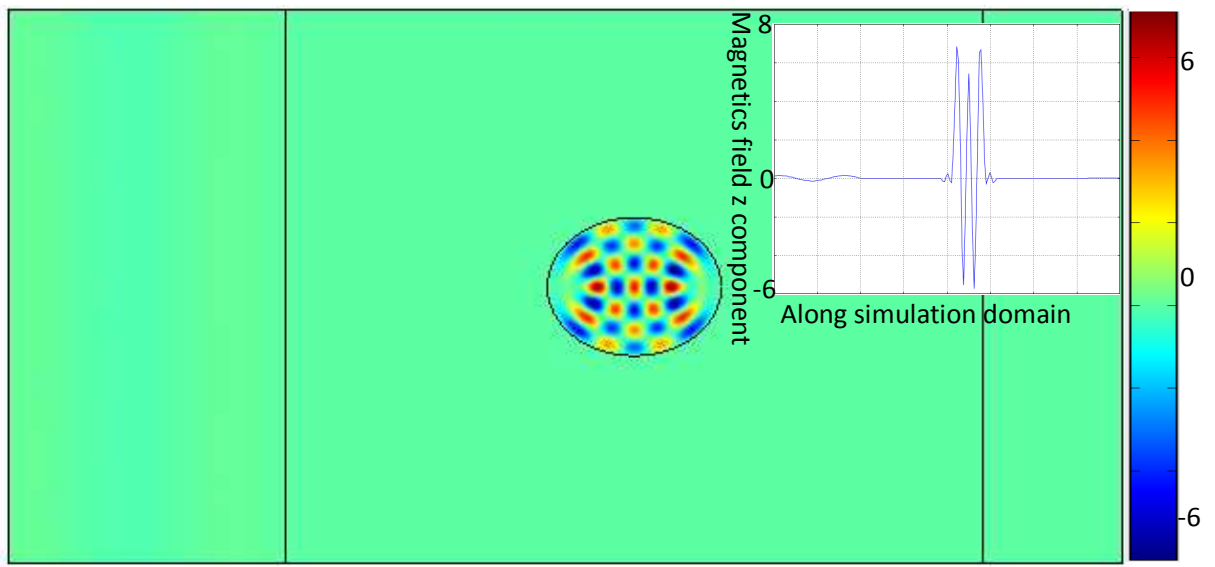

(b)

**Fig 5**

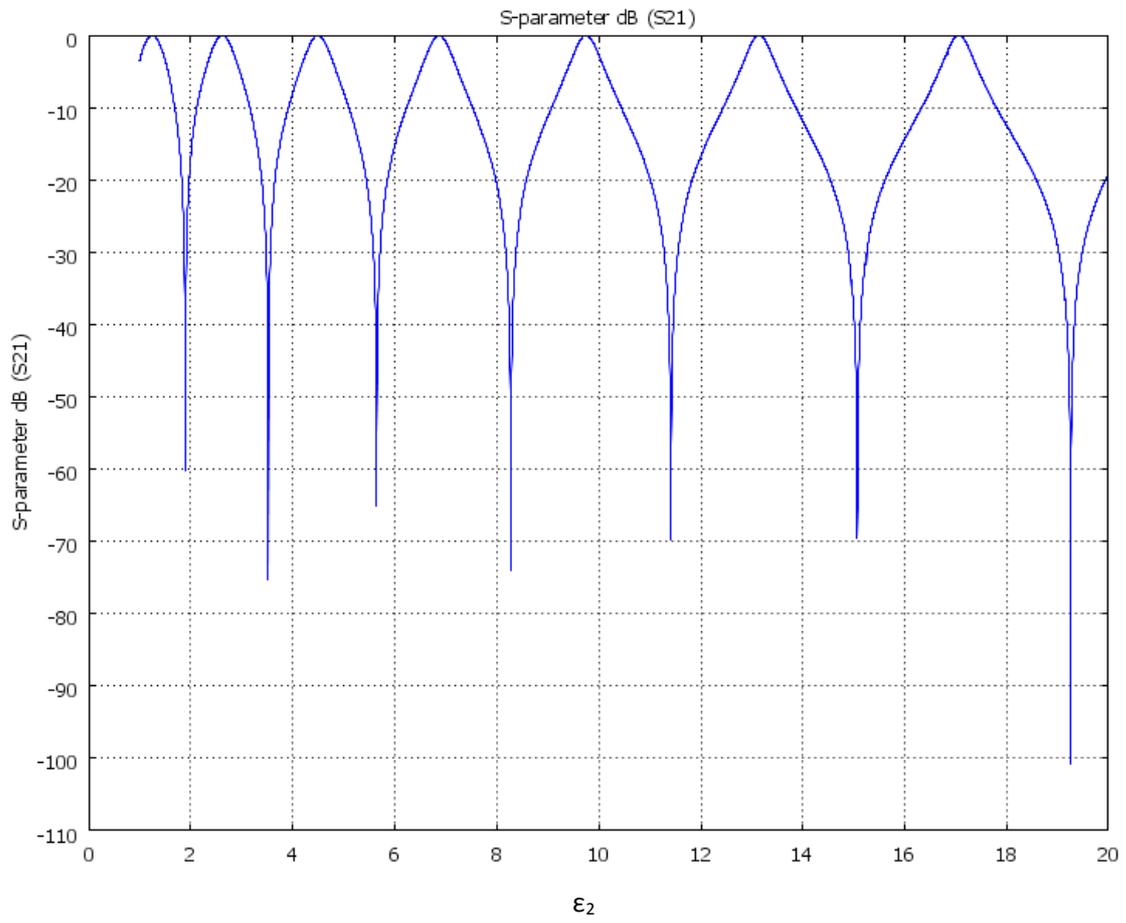

(a)

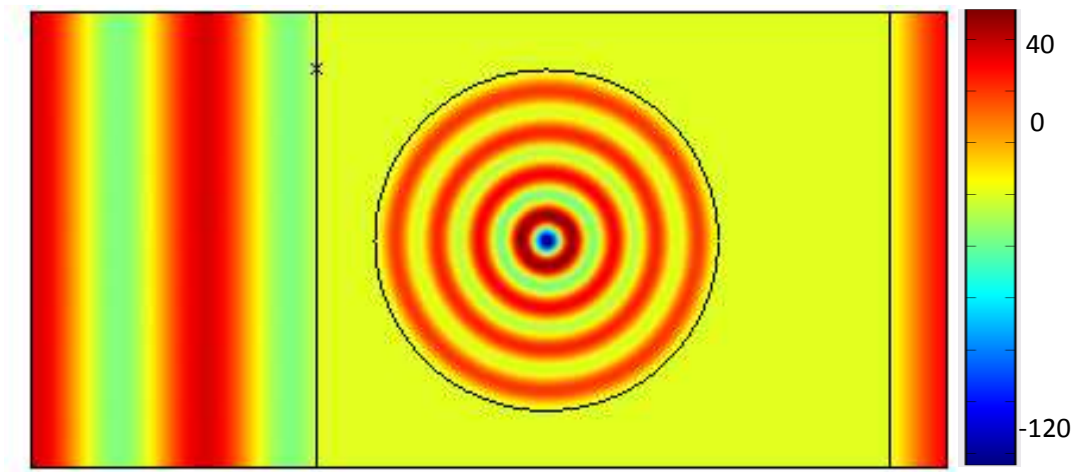

(b)